\documentclass[11pt,letterpaper]{article}
\usepackage{multicol}
\usepackage{latexsym}
\usepackage{amsmath}
\usepackage{latexsym,amssymb}
\usepackage{graphicx}
\usepackage{bm}
\usepackage{enumerate}
\usepackage{dsfont}
\usepackage{bbm}
\newcommand{\HRule}{\rule{\linewidth}{0.5mm}}
\newcommand{\HRuletwo}{\rule{0.40\linewidth}{0.25mm}}

\newcommand{\tinymath}[1]{\mbox{\tiny{#1}}}

\newcommand{\pb}{\mathcal{E}_{\tinymath{II}}}
\newcommand{\sigm}[4]{\Sigma^{#1\;\;\:#3}_{\;\:#2\;\;#4}}
\newcommand{\sinfun}{\sinh\left(\sqrt{\beta}\varrho\right)}
\newcommand{\cosfun}{\cosh\left(\sqrt{\beta}\varrho\right)}
\newcommand{\athree}{\mathcal{A}_{\tinymath{III}}}
\newcommand{\aone}{\mathcal{A}_{\tinymath{I}}}
\newcommand{\atwo}{\mathcal{A}_{\tinymath{II}}}
\newcommand{\ethree}{\mathcal{E}_{\tinymath{III}}}
\newcommand{\eone}{\mathcal{E}_{\tinymath{I}}}
\newcommand{\etwo}{\mathcal{E}_{\tinymath{II}}}
\newcommand{\bdot}{\dot{B}(T)}

\def \u1 {\textrm{U(1)}}

\begin{document}
\begin{multicols}{2}
\title{\bf{\LARGE A Note on the Symmetry Reduction of SU(2) on Horizons of Various Topologies.\newline\newline\newline\newline\newline\newline\newline\newline\newline}}
\author{{{\small Andrew DeBenedictis \footnote{adebened@sfu.ca}}} \\
\it{\footnotesize Department of Physics} \\
{\footnotesize  and} \\
\it{\footnotesize The Pacific Institute for the Mathematical Sciences} \\
\it{\footnotesize Simon Fraser University}\\
\it{\footnotesize Burnaby, BC, V5A 1S6, Canada} \\ \HRuletwo
 \and
{\small Steve Kloster \footnote{stevek@sfu.ca}} \\
\it{\footnotesize Centre for Experimental and Constructive Mathematics} \\
\it{\footnotesize Simon Fraser University}\\
\it{\footnotesize Burnaby, BC, V5A 1S6, Canada} \\ \HRuletwo
\and
{\small Johan Brannlund \footnote{johan\_brannlund@cbu.ca\,\hspace{-0.3mm},\,\hspace{0.3mm}johanb@mathstat.dal.ca}} \\
\it{\footnotesize Department of Mathematics, Physics and Geology} \\ 
\it{\footnotesize Cape Breton University} \\
\it{\footnotesize Sydney, NS, B1P 6L2, Canada}\\ \HRule 
}
\date{{March 22, 2011}}
\maketitle
\end{multicols}

\begin{abstract}
\noindent It is known that the SU(2) degrees of freedom manifest in the description of the gravitational field in loop quantum gravity are generally reduced to U(1) degrees of freedom on an $S^2$ isolated horizon. General relativity also allows black holes with planar, toroidal, or higher genus topology for their horizons. These solutions also meet the criteria for an isolated horizon, save for the topological criterion, which is not crucial. We discuss the relevant corresponding symmetry reduction for  black holes of various topologies (genus $0$ and $\geq 2$) here and discuss its ramifications to black hole entropy within the loop quantum gravity paradigm. Quantities relevant to the horizon theory are calculated explicitly using a generalized ansatz for the connection and densitized triad as well as utilizing a general metric admitting hyperbolic sub-spaces. In all scenarios, the internal symmetry may be reduced to combinations of U(1). 
\end{abstract}

\vspace{3mm}

\section{Introduction}
Loop quantum gravity is a candidate theory for quantum gravity that attempts to quantize general relativity in a diffeomorphism invariant way. (See, for example, \cite{ref:rovlrr}, \cite{ref:perez}, \cite{ref:rovbook}, \cite{ref:thiembook} and references therein.) One issue that is often cited as one that should be addressed by a theory of quantum gravity is that of the source of black hole entropy. Some believe that the source of this entropy is purely gravitational in nature, and that counting the number of gravitational quantum states attributable to the black hole should yield a measure of its entropy, and should agree with the $A/4$ law at lowest order. In loop quantum gravity, the gravitational field can be described by an SU(2) spin-network. Such a network has edges and vertices, and these give rise to a quantum geometry where the vertices may be associated with volume elements and the edges with ``fluxes of area''. This yields a natural way to associate quantum states with a black hole horizon. The spin network, when puncturing a surface, $\mathcal{S}$ endows the surface with an amount of area given by the eigenvalue \footnote{We are making an assumption here regarding how the spin-network pierces the surface $\mathcal{S}$. The general case yields eigenvalues which are slightly more complicated than (\ref{eq:aevals}).}
 \begin{equation}
 \Delta\mathcal{S} =8\pi\gamma\, \sqrt{j_{p}(j_{p}+1)} \, , \label{eq:aevals}
\end{equation}
where $\Delta\mathcal{S}$ denotes some surface element of $\mathcal{S}$, $\gamma$ is the Immirzi parameter, $p$ denotes which puncture is
under consideration, and $j_{p}$ can take on half-integer values which represents the spin carried by the puncture. The total area of $\mathcal{S}$ is given by adding up all the area eigenvalues contributed by all of the punctures on the surface. It is very interesting that the structure of space is naturally discrete in this theory.

The entropy of a black hole is then normally calculated as follows: One fixes the area of the black hole event horizon under consideration within some narrow range $(a_{0}-\epsilon)<a_{0}<(a_{0}+\epsilon)$. One then counts the number of ways spin-networks may puncture the surface, and yield a total area  within the allowed range. The logarithm of this number yields the entropy which, from (\ref{eq:aevals}), will involve the Barbero-Immirzi parameter $\gamma$. By setting the calculated entropy to $a_{0}/4$, one gets a value for $\gamma$. Since $\gamma$ plays a pivotal role in the theory, determining its value is of great importance. It should be noted that even with the $\gamma$ ambiguity, careful calculations have shown that the entropy using this technique is indeed linearly proportional to the area of the black hole \cite{ref:rovent}, \cite{ref:krasent}, \cite{ref:entrev6}, \cite{ref:numthy}. The degeneracy spectrum of black holes and its relation to the entropy has also recently been studied in \cite{ref:degen}. A calculation of the entropy in the SU(2) formalism incorporating quantum group corrections has recently been conducted in \cite{ref:enpp_new}.

From a phase-space point of view, what are known as \emph{isolated horizons} have been studied in the pioneering work of \cite{ref:ACK} and these isolated horizons have been used to study the entropy problem \cite{ref:ABK}, \cite{ref:ABCK}. It is found that on the isolated horizon (inner boundary) the SU(2) theory produces a Chern-Simons theory, which in turn reduces to only U(1) true degrees of freedom \cite{ref:ABK}, and this has ramifications for the sub-leading correction coefficient (for example, see \cite{ref:Gour}, \cite{ref:ghoshmitra}, and \cite{ref:combin}, the last reference utilising a combinatoric approach). There has been some ambiguity regarding this reduction. Since the original work several very interesting clarifying studies have emerged \cite{ref:KandM98}, \cite{ref:KandM00}. Specifically,  Engle, Noui and Perez \cite{ref:ENP} have studied the problem from a purely SU(2) perspective, which more easily reveals the connection between loop quantum gravity and the boundary theory that it produces on the horizon. More details on their work is provided in \cite{ref:ENPP}. In another series of interesting papers, by Basu, Kaul and Majumdar, they show that the U(1) formulation is equivalent to the SU(2) formulation subject to several natural constraints on the solder forms \cite{ref:BKandM}, \cite{ref:KandM}. The authors have also previously studied the problem by analogy with the SU(2) Wess-Zumino model \cite{ref:KandM98}, \cite{ref:KandM00}. As well, by studying the laws of black hole mechanics using weakly isolated horizons, the topological theory on the boundary of the black hole is a U(1) Chern-Simons theory \cite{ref:chatterjee2}. A U(1) result for spherical horizons has also been acheived by further relaxation of the horizon conditions, indicating that the U(1) nature is rather natural on black hole horizons \cite{ref:basu2}.  

Much of the work thus far has been done utilizing spherical topology; arguably the most physically relevant. However, general relativity admits horizons with other topologies such as planar, toroidal and higher genus topologies \cite{ref:rg1}-\cite{ref:mena}. Although these topologies may not be as physically relevant as their spherical counter-parts, there is good reason to study them. For example, quantum gravity is an arena that has very little experimental guidance. In regard of this, one has to resort to considerations of what one \emph{expects} from a theory of quantum gravity. Black hole entropy and the resolution of the classical singularities may be several desireable criteria for a viable quantum theory. One also wishes the theory to be self consistent in some way. That is, it should be able to produce the correct entropy and remove the singularity for all types of black holes found in the classical theory. In this vein, the exotic topologies have been studied in loop quantum gravity in \cite{ref:our_ent} and \cite{ref:our_sing}. In these studies the $A/4$ law was reproduced, as expected, and the sub-leading contribution was found to be genus dependent, which turns out to be in agreement with studies of higher genus black hole entropy utilizing non-loop quantum gravity techniques \cite{ref:vanzo}, \cite{ref:mannsol}, \cite{ref:liko}. Singularities were also resolved in a mini-superspace context under similar assumptions as with spherical black holes \cite{ref:our_sing}. 

In this note we wish to elaborate on the boundary structure of the higher genus horizons (although the treatment here is general enough to also cover the $g=0$ spherical black holes as well), especially regarding the issue of symmetry reduction on the horizon and its relation to the sub-leading order contribution to the entropy. In section 2 we briefly outline the conditions of an isolated horizon. In section 3 we discuss the boundary theory in the context of black holes of various topologies, discussing the role played by the internal symmetry group. We also provide a specific calculation to elucidate the arguments. Finally, we issue some concluding remarks in section 4. We use the notation that indices $i,\,j,\,k...$ etc. denote tetrad and SU(2) components and span the list $1,\,2,\,3$, whereas indices $a,\,b,\,c...$ etc. denote spatial components (not space-time, as we are explicitly using a 3+1 decomposition) and span the list $R,\,\varrho,\,\phi$. Greek indices span the full space-time.

\section{A brief review of isolated horizons}
The isolated horizon framework was originally developed by Ashtekar, Beetle and Fairhurst \cite{ref:ABF} based on earlier works by Hayward \cite{ref:hayhor}.
The minimal definitions of an isolated horizon most commonly found in the literature may be summarized as follows:
\begin{enumerate}[i.]
 \item The isolated horizon sub-manifold, denoted by $\Delta$, is topologically $S^{2}\times\mathbb{R}$ and is null. This topological restriction can be relaxed, and many results are insensitive to topology. However, allowing the surface to have a richer topology does lead to some interesting new results \cite{ref:our_ent}, \cite{ref:our_sing}.
\item The $\mathbb{R}$ sub-sector can be mapped to a null normal, denoted by $\ell^{\alpha}$. We will assume $\ell^{\alpha}$ is future pointing. Furthermore, on $\Delta$, $\ell^{\alpha}$ (and therefore any vector related to $\ell^{\alpha}$ by a constant re-scaling) possesses zero expansion. A second null vector on $\Delta$, $n^{\alpha}$, defined from the condition $\ell^{\alpha}n_{\alpha}=-1$, has negative expansion.
\item The equations of motion on $\Delta$ hold. Also, the flux vector, $-T^{\mu}_{\; \alpha}\ell^{\alpha}$, on $\Delta$ is
future-causal ($T^{\mu}_{\;\nu}$ being the stress-energy tensor of any matter fields present).
\end{enumerate}

\begin{figure}[ht]
\begin{center}
\includegraphics[bb=130 230 575 580, clip, scale=0.5, keepaspectratio=true]{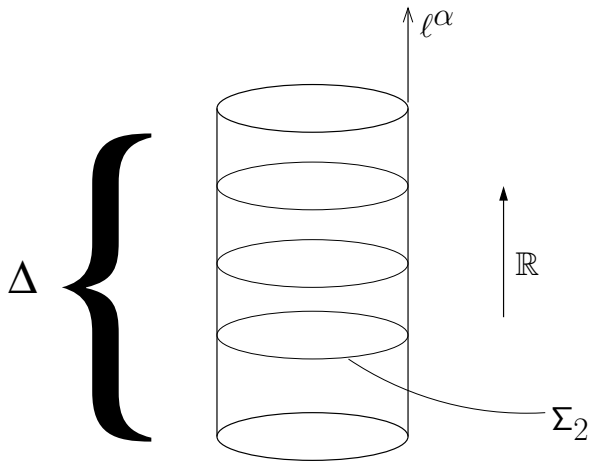}
\caption{{\small A schematic of an isolated horizon. The isolated horizon sub-manifold, $\Delta$, may be foliated by 2-surfaces of various topologies, $\mathsf{\Sigma}_{2}$.}}
\label{fig:IH}
\end{center}
\end{figure}

It should be stressed that these conditions are usually enforced \emph{only on} $\Delta$. Furthermore, $\Delta$ can be ``sliced'' into preferred foliations, which we denote as $\mathsf{\Sigma}_{2}$, and which are transverse to $\ell^{\alpha}$. A schematic is provided in figure~\ref{fig:IH}.

\section{Horizons of various topologies as isolated horizons and boundary conditions}
General relativity also admits solutions to the field equations representing asymptotically anti-de Sitter cylindrical, toroidal, and higher genus black holes. Such black holes, described below, are represented by asymptotically anti-de Sitter metrics with various 2-space symmetries (spherical, flat toroidal, and higher-genus hyperbolic). 

\subsection{From a general ansatz for $A$ and $E$}
For studies in the Ashtekar variables, we require an ansatz for a connection,  $A^{i}_{\;a}\,$, as well as a densitized triad, $E^{a}_{\;i}\,$, which is capable of accommodating the symmetries under study. We utilize the following for this:
\begin{subequations}
 \begin{align}
A=&\athree\,\tau_{1}\,dR + \left(\aone \tau_{2} + \atwo \tau_{3}\right)d\varrho +\left( \atwo\tau_{2} - \aone\tau_{3}\right)\sqrt{c}\sinh(\sqrt{\beta}\varrho)\,d\phi \nonumber \\
& -\sqrt{c}\sqrt{\beta}\cosh(\sqrt{\beta}\varrho)\,\tau_{1}\,d\phi\,, \label{eq:hyperA} \\
E=& -\mathcal{E}_{\tinymath{III}}  \sqrt{c}\sinh(\sqrt{\beta}\varrho)
\,\tau^{1}\,{\partial_R} - \left(\mathcal{E}_{\tinymath{I}}\tau^{2} +\mathcal{E}_{\tinymath{II}}\tau^{3}\right)\sqrt{c}\sinh(\sqrt{\beta}\varrho)\,{\partial_\varrho} \nonumber \\
& +\left(\mathcal{E}_{\tinymath{I}}\tau^{3}-\mathcal{E}_{\tinymath{II}}\tau^{2}\right)\,{\partial_\phi}\,, \label{eq:hyperE} 
 \end{align}
\end{subequations}
with $0 < \phi \leq 2\pi$ and where the functions $\mathcal{A}_{\cdot\cdot}$ and $\mathcal{E}_{\cdot\cdot}$ may be functions of the ``time'' coordinate, $T$, only. This ansatz is a generalization of a modification of Witten's spherically symmetric ansatz \cite{ref:wittenA}. It was shown in \cite{ref:our_sing} that (\ref{eq:hyperA}) and (\ref{eq:hyperE}) are sufficient to yield the spherical as well as the non-rotating higher genus black holes of general relativity. The cases are as follows:\\
i) $\beta=-1$, $c=-1$: In this case $(\varrho,\,\phi)$ sub-manifolds are spheres. \\
ii) $\beta=0$, $\underset{\beta\rightarrow 0}\lim\,c=\frac{1}{\beta}$: In this case $(\varrho,\,\phi)$ sub-manifolds are tori. Event horizon surfaces for this case are intrinsically flat. \\
iii) $\beta=1$, $c=1$: In this case $(\varrho,\,\phi)$ sub-manifolds are surfaces of constant negative curvature of genus $g > 1$, depending on the identifications chosen. Such surfaces may be compact or non-compact \cite{ref:complexteich}, \cite{ref:geo3}. \\

It turns out that the torus case ($g=1$) is exceptional due to the fact that the pull-back of the SU(2) connection on to the $R=\mbox{const.}$ two-torus is constant and can be gauged to zero. This is not an issue in the higher genus cases.

It should be noted that in the $g\neq 0$ cases the coordinate $\varrho$ is periodic for the construction to work. This also enforces the uniqueness of relevant quantities under large translations. 

We start by imposing the Gauss constraint to eliminate excess gauge rotational freedom. In the system above, the Gauss constraint reads: 
\begin{equation}
G_{i}:=\partial_{a}E^{a}_{\;i} +\epsilon_{ij}^{\;\;\;k}A^{j}_{\;a}E^{a}_{\;k}=0\,, 
\end{equation}
which in the scenarios studied here yields only one non-trivial condition:
\begin{equation}
 2\sqrt{c}\sinh\left(\sqrt{\beta}\varrho\right)\left[\atwo \mathcal{E}_{\tinymath{I}}-\aone\pb\right]=0\,. \label{eq:hypergc}
\end{equation}
To satisfy (\ref{eq:hypergc}) we set $\atwo=\etwo=0$. This amounts only to a partial gauge fixing, eliminating redundant degrees of freedom, and therefore does not affect the physical conclusions.

The field strength tensor will also be required later and may be calculated via:
\begin{equation}
 F^{i}_{\;ab}=\partial_{a}A^{i}_{\;b}-\partial_{b}A^{i}_{\;a}+\epsilon^{i}_{\;jk}A^{j}_{\;a}A^{k}_{\;b} \label{eq:Fdef}
\end{equation}
This yields (subject to the fixed Gauss constraint):
\begin{subequations}\allowdisplaybreaks
\begin{align}
F^{1}_{\;\varrho\phi}=&-\left(\aone^{2}+\beta\right)\sqrt{c}\,\sinfun\,, \label{eq:genF1}\\
F^{2}_{\;R\phi}=&\athree\aone\sqrt{c}\,\sinfun\,, \label{eq:genF2}\\
F^{3}_{\;R\varrho}=&\athree\aone\,,\label{eq:genF3}
\end{align}
\end{subequations}

In the variables utilized here, the boundary term on $\Delta$ (often called the inner boundary to distinguish it from infinity or other horizons, such as the de Sitter horizon) that arises in varying the gravitational action with respect to the connection takes the form:
\begin{equation}
\delta I_{\Delta}=-\frac{1}{8\pi l_{p}}\int_{\Delta} \mbox{Tr}\left[\Sigma \wedge \delta A\right]\,, \label{eq:bt}  
\end{equation}
where $l_{p}$ denotes the Planck length and $\Sigma$ is often called the `solder form'', which can be constructed from the triad and the metric-independent Levi-Civita, $\eta^{abc}$. Specifically, 
\begin{equation}
 E^{a}_{\;i} =\frac{1}{2}\eta^{abc}\,\sigm{j}{b}{k}{c}\,\epsilon_{ijk}
\end{equation}
yields:
\begin{align}
\Sigma:=\sigm{i}{a}{j}{b}\, \tau_{i}\, dx^{a} \tau_{j}\, dx^{b}= &\frac{1}{2}\,\eone\, \tau^{3}\,  dR\wedge d\varrho + \frac{1}{2}\,\eone\,\sqrt{c}\sinfun\,\tau^{2}\,   dR\wedge d\phi \nonumber \\
& -\frac{1}{2}\,\ethree\sqrt{c}\sinfun\, \tau^{1}\, d\varrho\wedge d\phi\,. \label{eq:gensolder}
\end{align}

Now, the fact that the vector $\ell^{\alpha}$ is null dictates that, on $\Delta$, $\eone\widehat{=}0$, where $\widehat{=}$ denotes that the equality only must hold on $\Delta$. Therefore, from (\ref{eq:gensolder}) only the $\tau^{1}$ component of $\Sigma$ survives. The equations of motion dictate that $\aone$ must also vanish on $\Delta$ and that only components $A^{1}_{\;a}$ are therefore non-zero on the inner boundary, indicating that the theory has a U(1) content. Furthermore, the zero expansion condition dictates that $\ethree$ must be constant valued on $\Delta$ and therefore, by comparing (\ref{eq:genF1}) to (\ref{eq:gensolder}) on $\Delta$,  allows us to re-write (\ref{eq:bt}) as
\begin{equation}
\delta I_{\Delta}=\frac{K_{0}}{8\pi l_{p}}\int_{\Delta} \mbox{Tr}\left[F\wedge \delta A\right]\, \label{eq:Fbt}
\end{equation}
which, from (\ref{eq:Fdef}) yields:
\begin{equation}
\delta I_{\Delta}=\frac{K^{\prime}_{0}}{8\pi l_{p}}\,\, \delta \hspace{-1mm}\int_{\Delta}\mbox{Tr}\left[A\wedge dA +\frac{2}{3} A\wedge A \wedge A\right]\,. \label{eq:CSbt}
\end{equation}
Here, $K_{0}$ and $K^{\prime}_{0}$ are constants of proportionality related to the Chern-number of the theory.

Note that on $\Delta$, due to the fact that $\aone$ and $\atwo$ are both zero, we may write the $A$ connection as a U(1) connection; $A\rightarrow W:= \athree \tau_{1}\,dR -\tau_{1}\sqrt{c}\sqrt{\beta}\cosh(\sqrt{\beta}\varrho)\,d\phi$, and the previous expression reduces to
\begin{equation}
\delta I_{\Delta}=\frac{K^{\prime}}{8\pi l_{p}}\,\, \delta \hspace{-1mm}\int_{\Delta}\mbox{Tr}\left[W\wedge dW\right] = -\frac{K^{\prime}}{8\pi l_{p}}\,\, \delta \hspace{-1mm}\int_{\Delta}\mbox{Tr}\left[\athree\sqrt{c}\beta\,\sinfun\right]\,, \label{eq:u1bt}
\end{equation}
yielding a boundary action for a U(1) theory. Note that, as mentioned previously, the $g=1$ torus case, $T^{2}=S^{1}\times S^{1}$, is exceptional from the other cases and yields a trivial theory. (However, certain results, such as entropy, pertaining to higher genus black holes may be analytically extended to encompass the $g=1$ scenario \cite{ref:our_ent}.) Regarding the triviality of the $g=1$ scenario, it is possible to rule out many possibilities for the total space ${\mathsf{E}}$ for a bundle
with base $S^1 \times S^1$ and fiber $\u1 $, using techniques from algebraic
topology. (For the necessary background for these techniques, see \cite{hatcher}.) To do this, note that a bundle ${\mathsf{E}} \to B$ with fiber $\mathcal{F}$ gives rise to a long exact sequence

\begin{displaymath}
  ... \to \pi_n(\mathcal{F}) \to \pi_n({\mathsf{E}}) \to \pi_n(B) \to \pi_{n-1}(\mathcal{F}) \to \pi_{n-1}({\mathsf{E}})
\to \pi_{n-1}(B) \to ... \to \pi_0(B) \to 0\,.
\end{displaymath}

In this case, $B=S^1 \times S^1$ and $\mathcal{F}=\textrm{U(1)}$, which gives us

\begin{displaymath}
  \pi_3(\u1 ) \to \pi_3({{\mathsf{E}}}) \to \pi_3(S^1 \times S^1)\,.
\end{displaymath}

Since $  \pi_3(\u1 )=0$, and $\pi_3(S^1 \times S^1)=0$, we find the
exact sequence $0 \to \pi_3({\mathsf{E}}) \to 0$. Exactness of this sequence implies that
$\pi_3({\mathsf{E}})=0$, which rules out, among many other things, ${\mathsf{E}}=S^3$ (to which $SU(2)$ is diffeomorphic). One
remaining possibility is ${\mathsf{E}}=T^3$.

We can actually get somewhat more of a handle on the structure of the possibly allowed
bundles by using a slightly different point of view.
$\u1 $ bundles over $S^1 \times S^1$ are classified by 
the classifying space $[S^1 \times S^1,\mathbb{C}P^\infty]$, which consists
of homotopy classes of maps from $S^1 \times S^1$ to
$\mathbb{C}P^\infty$ (infinite-dimensional complex projective space). Since
the latter can be thought of as the Eilenberg-MacLane space $K(\mathbb{Z},2)$, it
follows that the classifying space is the second singular cohomology 
$H^2(S^1 \times S^1,\mathbb{Z})$. It follows (for instance from Poincar\'e duality)
that  $H^2(S^1 \times S^1,\mathbb{Z})=\mathbb{Z}$, so there exists an integer's worth of bundles with base $S^1 \times S^1$ and fiber $\u1 $, none of which correspond to a total space of ${\mathsf{E}}=S^3$.

It should be noted that the above comments apply only on ${\mathsf{\Sigma}}_{2}=T^{2}$ ($g=1$) and they do \emph{not} imply that one cannot have a U(1) theory or SU(2) theory on a torus, but they do indicate that one cannot possess a U(1) theory whose total space is $S^{3}$, meaning that a U(1) theory on $T^2$  cannot come from an SU(2) reduction.  Arguments similar to the above can be used to show that U(1) is the natural fiber over $B=S^{2}$. As well for the sphere, one may consider the Bianchi identity $dF+A\wedge F-F\wedge A=0$. If the wedge product terms vanish, the identity then implies that $F$ is exact on $S^{2}$. Therefore, the field-strength reduces to the Abelian version on $S^{2}$ (after possibly further gauge transformations).
This of course is not necessarily true for the higher genera, and the fact that the surface obeys the horizon conditions is crucial for the reduction of the boundary action to a U(1) theory, as shown above. (More specifically, a topological basis can be constucted on the surface that is a connected sum of $g$ copies of U(1)$\times$U(1)).

\subsection{Specific example from a metric}
In general relativity, a reasonably general metric capable of describing black holes of various topologies is provided by
\begin{equation}
ds^2=-\mathcal{B}(t,\,r)\,dt^{2} + \mathcal{C}(t,\,r)\,dr^{2} +r^2\left(d\varrho^{2} + c\,\sinh^{2}(\sqrt{\beta}\,\varrho)\,d\phi^{2}\right), \label{eq:smmetric}
\end{equation}
 Here,  $c$ and $\beta$ are constants adapted to determine the topology of $\varrho,\phi$ sub-surfaces as described earlier. 
An event horizon exists when $\mathcal{B}(t,\,r)=0$. We wish to briefly show here that this horizon satisfies the relevant properties of isolated horizons. Before continuing, we re-write the line element of this space-time in a coordinate chart more suitable for the interior of the black hole (sometimes called the ``$T$-domain''). This will prove to be useful for some of the subsequent analysis. In the interior chart, line element (\ref{eq:smmetric}) may be recast as
\begin{equation}
 ds^{2}= -C(R,\,T)\,dT^{2} + B(R,\,T)\,dR^{2} + T^{2}\left(d\varrho^{2} + c\sinh^{2}(\sqrt{\beta}\,\varrho)\,d\phi^{2}\right), \label{eq:Tmetric}
\end{equation}
with coordinate ranges:
\begin{equation}
 0 < T \leq T_{\Delta}\,,\;\;\; R\in \mathbb{R} \,,\;\;\; 0 < \varrho < \varrho_{1} \,,\;\;\; 0 \leq \phi < 2\pi\,. \nonumber
\end{equation}
The $T$-domain version of the metric proves to be more useful in this calculation as, on the horizon, the $R$ direction is coincident with the direction of the null vector $\ell^{\alpha}$ used in the definition of isolated horizons. 

The condition that the ($\Lambda$) vacuum field equations hold on $\Delta$ imply both the conditions $C(R,\,T)=C(T)$ and that $C(T)\propto 1/B(T)$. (Further restrictions from the field equations will not be needed.) Furthermore, the condition that $\ell^\alpha$ is null dictates that $B(T)\widehat{=}0$ on $\Delta$. (The symbol $\widehat{=}$ is often used in the literature to denote that an equality need only hold on $\Delta$.)

The metric given by the line element in (\ref{eq:Tmetric}) will be useful in providing an explicit check of the calculations carried out in the previous sub-section. Such a check has also been employed in \cite{ref:KandM} where the explicit form of the Schwarzschild metric was used to illustrate that the sub-leading correction of the entropy for $S^{2}$ isolated horizons is indeed $-\frac{3}{2}$, although a four-dimensional approach was utilized there whereas a 3-space approach, adapted to the 3+1 Hamiltonian formalism, will be utilized here.

In the 3+1 formalism, which is often utilized in the Hamiltonian approach to quantum gravity, the 3-metric $q_{ab}$ is used to calculate many of the relevant quantities,
\begin{equation}
d\sigma^{2}=q_{ab}\,dx^{a}dx^{b}= B(T)\,dR^{2} + T^{2}\left(d\varrho^{2} + c\,\sinh^{2}(\sqrt{\beta}\,\varrho)\,d\phi^{2}\right)\:. \label{eq:3Tmetric}
\end{equation}
Although in this chart the metric is partially degenerate on the horizon, all relevant quantities will turn out to be insensitive to this degeneracy and in fact are \emph{continuous and non-pathological} across the horizon. (It should be noted that ``outside'' of the horizon, quantities such as $\sqrt{B(T)}$ should be replaced by $\sqrt{|B(T)|}$, but we are approaching the horizon from the interior and therefore omit the absolute value.)

The 3-metric in (\ref{eq:3Tmetric}) admits the natural orthonormal tetrad\footnote{The orientation of the triad here is chosen to be compatible with the coordinate system in (\ref{eq:hyperA}) and (\ref{eq:hyperE}).}:
\begin{equation}
 e^{i}_{\;\:a}=\sqrt{B(T)}\,\delta^{i}_{\;1}\,\delta^{R}_{\;a}+ T\,\delta^{i}_{\;2}\,\delta^{\varrho}_{\;a}-\sqrt{c}\,T\,\sinfun\, \delta^{i}_{\;3}\,\delta^{\phi}_{\;a}\,,
\end{equation}
which yields the following densitized triad via $E^{a}_{\;i}=\frac{1}{2}\epsilon^{abc}\epsilon_{ijk}e^{j}_{\;b}e^{k}_{\;c}\;$:
\begin{equation}
E = -T^{2}\sqrt{c}\sinfun\,\tau^{1}\,\partial_{R} - T \sqrt{c} \sinfun\,\sqrt{B(T)}\,\,\tau^{2}\,\partial_{\varrho}+ T\sqrt{B(T)}\,\tau^{3}\,\partial_{\phi}\,.   \label{eq:dtriad}
\end{equation}

The conjugate configuration variable to the densitized triad is the Barbero-Immirzi connection, given by 
\begin{equation}
A^{i}_{\;a}:=\Gamma^{i}_{\;a}+\gamma K^{i}_{\;a}. \label{eq:biconnection}
\end{equation} 
Here, $\Gamma^{i}_{\;a}$ is the ``fiducial'' spin connection whose associated derivative annihilates the triad via:
\begin{equation}
 \partial_{[a}e^{i}_{\;b]}+\epsilon^{i}_{\;jk}\Gamma^{j}_{\;[a}e^{k}_{\;b]}=0\,, \label{eq:cartan}
\end{equation}
and which is explicitly provided by:
\begin{equation}
 \Gamma^{i}_{\;a}=-\frac{1}{2}\epsilon^{ij}_{\;\;\;k}e^{\;\;b}_{j}\left[\partial_{a}e^{k}_{\;b}-\partial_{b}e^{k}_{\;a} +\delta^{kl}\delta_{mn}e_{l}^{\;c}\,e^{m}_{\;a}\partial_{b}e^{n}_{\;c}\right]\,. \label{eq:Gam}
\end{equation}
Finally, $K^{i}_{\;a}$ is related to the extrinsic curvature, $K_{ab}$, of a $T=T_{0}$ surface via
\begin{equation}
 K^{i}_{\;a}:=\frac{1}{\sqrt{\mbox{det}(E)}}\delta^{ij} K_{ab}E^{b}_{\;j}\,. \label{eq:modext}
\end{equation}

Using (\ref{eq:Gam}) and (\ref{eq:modext}) in (\ref{eq:biconnection}) we explicitly calculate the connection as
\begin{align}
A=&-\frac{\gamma}{2}\, \bdot\,\tau_{1}\,dR -\sqrt{c}\sqrt{\beta}\cosfun\,\tau_{1}\,d\phi 
- \gamma\sqrt{B(T)}\,\tau_{2}\,d\varrho \nonumber \\
& + \gamma\sqrt{c}\sinfun\,\sqrt{B(T)}\,\tau_{3}\,d\phi\,, \label{eq:connection}
\end{align}
where the overdot denotes differentiation with respect to $T$. Note that the above densitized triad and connection are compatible with the general ansatz (\ref{eq:hyperA}), (\ref{eq:hyperE}) in the case when $\atwo=\etwo=0$. By comparison with (\ref{eq:hypergc}) this is perhaps not surprising, since by choosing the coordinate system (\ref{eq:3Tmetric}) we have already partially gauge fixed the system. This is equivalent to the statement that the Gauss constraint is satisfied. 

With all the above we can now calculate the remaining quantities required for the boundary theory (\ref{eq:bt}); namely the SU(2) field strength tensor and solder forms. The field tensor components are provided by
\begin{subequations}\allowdisplaybreaks
\begin{align}
F^{1}_{\;\varrho\phi}=&-\left[\beta+\gamma^{2}B(T)\right]\sqrt{c}\sinfun\,,  \\
F^{2}_{\;R\phi}=&\frac{\gamma^{2}}{2}\sqrt{B(T)}\, \bdot\,\sqrt{c}\sinfun\,, \\
F^{3}_{\;R\varrho}=&\frac{\gamma^{2}}{2}\sqrt{B(T)}\, \bdot\,,
\end{align}
\end{subequations}
where we have not listed the components related via $F^{i}_{\;ab}\equiv -F^{i}_{\;ba}$.

Finally, the solder forms $\Sigma^{i\;\;j}_{\;a\;\;b}:=e^{i}_{\;[a}e^{j}_{\;b]}$ are calculated as (for brevity we again omit those related by anti-symmetries):
\begin{subequations}\allowdisplaybreaks
\begin{align}
 \sigm{1}{R}{2}{\varrho}=&\sigm{2}{\varrho}{1}{R}=\frac{T}{2}\sqrt{B(T)}\,, \label{eq:solder1}\\
 \sigm{1}{R}{3}{\phi}=&\sigm{3}{\phi}{1}{R}=-\frac{T}{2}\sqrt{B(T)}\,\sqrt{c}\sinfun\,, \label{eq:solder2} \\
 \sigm{2}{\varrho}{3}{\phi}=&\sigm{3}{\phi}{2}{\varrho}=-\frac{T^{2}}{2}\sqrt{c}\sinfun =: \Sigma^{(1)}_{\varrho\phi}\,. \label{eq:solder3}
\end{align}
\end{subequations}

Having constructed the relevant quantities we now consider their properties on the horizon itself. From the condition that $B(T)\widehat{=}0$ on the horizon, the quantities calculated reduce drastically on the horizon to:
\begin{subequations}
\begin{align}
E\,\widehat{=}\,&-T^{2}\sqrt{c}\sinfun\,\tau^{1} \partial_{R}\,, \label{eq:hordt} \\
A\,\widehat{=}\,&\left[-\frac{\gamma}{2}\dot{B}(T)\,dR - \sqrt{c}\sqrt{\beta}\cosfun\,d\phi\right]\tau_{1} \,, \label{eq:horcon} \\
F\,\widehat{=}\,&-\beta\sqrt{c}\,\sinfun\, \tau_{1}\,d\varrho\wedge d\phi \,, \label{eq:horF} \\
\Sigma\,\widehat{=}\,&-\frac{T^{2}}{2}\sqrt{c}\sinfun\,\tau^{1}\,d\varrho\wedge d\phi \,, \label{eq:horsig}
 \end{align}
\end{subequations}
 From these it can immediately be noted that the quantities, on the horizon, are U(1) valued, and therefore the theory governing their dynamics is also a U(1) theory. 

In \cite{ref:our_ent} it was shown that the Chern-level, $k$, of the theory for higher-genus scenarios is given by $k=\frac{a_{0}}{4\pi\gamma(g-1)}$ where $a_{0}$ is the (fixed) horizon area and $g$ is the genus of the horizon.
From (\ref{eq:hordt}-d), the following relationship therefore holds between the field-strength and the solder forms on the horizon:
\begin{equation}
 F^{1}_{\;\varrho\phi}\widehat{=}\frac{\beta \sqrt{c}\, \tilde{a}_{0}}{2\pi\sqrt{\beta}k\gamma(g-1)}\cdot\Sigma^{(1)}_{\varrho\phi}\,, \label{eq:fsigmarel}
\end{equation}
with $\tilde{a}_{0}:=\int_{\phi=0}^{2\pi}\left|\cosh\left[\sqrt{\beta}\,\varrho_{1}\hspace{-0.45mm}(\phi)\right]-1\right|\,d\phi$, which comes from the area integral \footnote{That is, in the general case,the compact surface has an upper-limit along some curve given by $\varrho=\varrho_{1}\hspace{-0.45mm}(\phi)$. In the spherical case, $\varrho_{1}\hspace{-0.45mm}(\phi)=\mbox{constant}=\pi$.}. Here, the ``$(1)$'' index denotes that this is the $\tau^{1}$ component of $\Sigma$. Furthermore, we have the following conditions on $\Delta$:
\begin{equation}
F^{1}_{\;R\varrho}\widehat{=}0\,,\;\;\;\mbox{and}\;\;\; F^{1}_{\;R\phi}\widehat{=}0\,. \label{eq:fzero}
\end{equation}
These conditions essentially boil down to those cited in \cite{ref:KandM} for the case of $S^{2}$ horizons but with a more complicated and genus dependent coupling coefficient. The complication is expected as the calculation here covers all cases. Again it can be seen that the $g=1$ case is pathalogical. 

At this stage it can be seen that the above system is equivalent to a gauge-fixed U(1) sub-group of an SU(2) Chern-Simons theory with sources. The sources can be interpreted to arise from considerations of the quantum theory, where the genus $g$ surface is replaced by a genus $g$ surface with punctures from the quantum gravitational spin-network, and the punctures act as source terms. A 2+1 dimensional SU(2) Chern-Simons theory with source possesses an action of the form
\begin{equation}
 {S}_{\mbox{\tiny{CS}}}=\mu_{0} \int_{\Delta}\mbox{Tr}\left[A\wedge dA+\frac{2}{3}A\wedge A \wedge A \right] +  \int_{\Delta} \mbox{Tr}\left[J\cdot A\right]\, \label{eq:csact}
\end{equation}
where $\mu_{0}$ is a constant related to the Chern level of the theory. Variation with respect to the potential $A$ yields the equations of motion
\begin{equation}
 \mu_{0} \eta^{abc}F^{i}_{\;bc}=J^{ai}\,. \label{eq:csEOM}
\end{equation}
Note that in the scenario studied here, from (\ref{eq:horcon}), the second term in the first integral of (\ref{eq:csact}) vanishes, thus producing the action for an \emph{Abelian} Chern-Simons theory. In this theory, the only non-trivial equation of motion is
\begin{equation}
F^{1}_{\;\varrho\phi}=\frac{1}{\mu_{0}}J^{R1}\,, \label{eq:eomsurvive}
\end{equation}
where we can identify the source term with the solder form $J^{R1}=\frac{\beta\sqrt{c}\tilde{a}_{0}}{8\pi^{2}\sqrt{\beta}\gamma(g-1)}\cdot\Sigma^{(1)}_{\;\varrho\phi}\,$ from (\ref{eq:fsigmarel}), the other two components of the source vanishing via the conditions (\ref{eq:fzero}). Therefore, classically, the boundary theory is indeed equivalent to a gauge-fixed ($F^{1}_{R\varrho}\widehat{=}0,\, F^{1}_{R\phi}\widehat{=}0$) U(1) Chern-Simons theory with U(1) source current $J$. This is analogous to the results obtained for the $S^{2}$ horizons in \cite{ref:ENPP} and \cite{ref:KandM}.

\begin{figure}[ht]
\begin{center}
\includegraphics[bb=0 0 713 135, clip, scale=0.5, keepaspectratio=true]{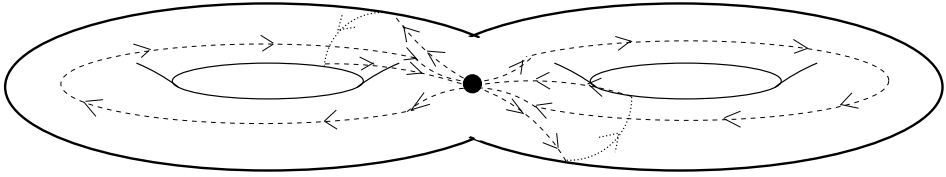}
\caption{{\small Half of the holonomy path on $\mathsf{\Sigma}_{2}$ for the genus 2 case. The other half of the path traverses the shown path but in the reverse direction.}}
\label{fig:holonomy}
\end{center}
\end{figure}

The difference between the $S^{2}$ case and the higher genus cases is the presence of non-trivial cycles on the higher genus surface, even in the absence of the punctures. This means that holonomy paths on the surface may be decomposed into a basis of holonomies along these cycles. The above results imply that, on a specific foliation $\mathsf{\Sigma}_{2}$ of $\Delta$, holonomies need to be considered in the $\phi$ direction. Half of such a path (for the range $0 < \phi < \pi$) for the $g=2$ case is shown in figure \ref{fig:holonomy} and this path is decomposable into non-contractible cycles of $\mathsf{\Sigma}_{2}$ even before the spin-network punctures are introduced in quantization. 

When the punctures from the gravitational spin-network are added, the genus cycles add a non-trivial relationship amongst the topological degrees of freedom of the spin-network. Namely, the following condition must be respected:
\begin{equation}
 \eta_{g+1}\cdot\eta_{g+2}\cdot ... \cdot \eta_{g+N}= \eta_{1}\gamma_{1}\eta_{1}^{-1}\gamma_{1}^{-1} \cdot ... \cdot \eta_{g}\gamma_{g}\eta_{g}^{-1}\gamma_{g}^{-1}. \label{eq:torusrelg}
\end{equation}
Here, $\eta_{g+1},\,...,\,\eta_{g+N}$ represent cycles around the $N$ punctures from the spin-network whereas $\eta_{1},\, ... , \eta_{g}$ and $\gamma_{1},\, ... , \gamma_{g}$ represent cycles around the poloidal and toroidal paths of the genus $g$ surface respectively. The symplectic structure to be quantized is then of the form
\begin{equation}
 \omega=\frac{k}{2\pi} \sum^{g+N-1}_{n=1} \left[ \delta \mathsf{A}_{n} \delta \mathsf{B}^{\prime}_{n} - \delta \mathsf{B}_{n} \delta \mathsf{A}^{\prime}_{n}\right]\,, \label{eq:symp}
\end{equation}
where $\delta \mathsf{A}_{n}$ and $\delta \mathsf{B}_{n}$ are U(1)-valued forms dual to the non-trivial cycles due to the surface  and the punctures, and their conjugate paths respectively. One then quantizes the symplectic structure (\ref{eq:symp}) subject to the constraint (\ref{eq:torusrelg}). It is the constraint (\ref{eq:torusrelg}) which gives rise to a genus dependent sub-leading coefficient to the entropy of the black hole \cite{ref:our_ent}. In a different context, Chern-Simons theories at the classical and quantum level have been studied at $g>0$ in, for example, \cite{ref:CShighgdunne}-\cite{ref:CShighggelca} and references therein.


\section{Concluding remarks}
In this note it has been shown how an explicitly SU(2) theory defined on a hyperbolic (for $g>1$) or spherical (for $g=0$) isolated horizon reduces to a topological theory of U(1) connections. This has been shown two ways; by utilizing a symmetry respecting connection and densitized triad directly, as well as by a method utilizing a metric capable of describing such black holes that arise in general relativity. In all cases the U(1) theory arises naturally from the SU(2) theory via a reduction, due to the space-time properties of the isolated horizon, to a U(1) sub-group of SU(2). Therefore, as with the $S^{2}$ horizons, the U(1) theory in the topologically non-trivial cases is simply a reduced SU(2) theory with the further constraints (\ref{eq:fzero}) and the physical contents of both the SU(2) and U(1) avenues of study are equivalent. The toroidal scenario is exceptional in that it yields a trivial theory.

\section*{Acknowledgments}
The authors are grateful to Ingemar Bengtsson of Stockholm University for helpful discussions regarding group symmetry reductions.

\linespread{0.6}
\bibliographystyle{unsrt}

\end{document}